# Validation of the Virtual Reality Neuroscience Questionnaire: Maximum Duration of Immersive Virtual Reality Sessions Without the Presence of Pertinent Adverse Symptomatology.

**Panagiotis Kourtesis[1,2,3,4], Simona Collina[3,4], Leonidas A.A. Doumas[2], and Sarah E. MacPherson[1,2]**

[1] Human Cognitive Neuroscience, Department of Psychology, University of Edinburgh, Edinburgh, United Kingdom

[2] Department of Psychology, University of Edinburgh, Edinburgh, United Kingdom

[3] Lab of Experimental Psychology, Suor Orsola Benincasa University of Naples, Naples, Italy

[4] Interdepartmental Centre for Planning and Research "Scienza Nuova", Suor Orsola Benincasa University of Naples, Naples, Italy

**\* Correspondence:**
Panagiotis Kourtesis
pkourtes@exseed.ed.ac.uk

## Abstract

There are major concerns about the suitability of immersive virtual reality (VR) systems (i.e., head-mounted display; HMD) to be implemented in research and clinical settings, because of the presence of nausea, dizziness, disorientation, fatigue, and instability (i.e., VR induced symptoms and effects; VRISE). Research suggests that the duration of a VR session modulates the presence and intensity of VRISE, but there are no suggestions regarding the appropriate maximum duration of VR sessions. The implementation of high-end VR HMDs in conjunction with ergonomic VR software seems to mitigate the presence of VRISE substantially.  However, a brief tool does not currently exist to appraise and report both the quality of software features and VRISE intensity quantitatively.

The Virtual Reality Neuroscience Questionnaire (VRNQ) was developed to assess the quality of VR software in terms of user experience, game mechanics, in-game assistance, and VRISE. Forty participants aged between 28 and 43 years were recruited (18 gamers and 22 non-gamers) for the study. They participated in 3 different VR sessions until they felt weary or discomfort and subsequently filled in the VRNQ.

Our results demonstrated that VRNQ is a valid tool for assessing VR software as it has good convergent, discriminant, and construct validity. The maximum duration of VR sessions should be between 55-70 minutes when the VR software meets or exceeds the parsimonious cut-offs of the VRNQ and the users are familiarized with the VR system. Also. the gaming experience does not seem to affect how long VR sessions should last. Also, while the quality of VR software substantially modulates the maximum duration of VR sessions, age and education do not. Finally, deeper immersion, better quality of graphics and sound, and more helpful in-game instructions and prompts were found to reduce VRISE intensity.

The VRNQ facilitates the brief assessment and reporting of the quality of VR software features and/or the intensity of VRISE, while its minimum and parsimonious cut-offs may appraise the suitability of VR software for implementation in research and clinical settings. The findings of this study contribute to the establishment of rigorous VR methods that are crucial for the viability of immersive VR as a research and clinical tool in cognitive neuroscience and neuropsychology.

## 1    Introduction

Immersive virtual reality (VR) has emerged as a novel tool for neuroscientific and neuropsychological research (Bohil *et al.*, 2011; Parsons, 2015; Parson *et al.*, 2018). Nevertheless, there are concerns pertinent to implementing VR in research and clinical settings, especially regarding the head-mounted display (HMD) systems (Sharples *et al.*, 2008; Bohil *et al.*, 2011; de Franca & Soares, 2017; Palmisano *et al.*, 2017). A primary concern is the presence of adverse physiological symptoms (i.e., nausea, dizziness, disorientation, fatigue, and postural instability), which are referred to as motion-sickness, cybersickness, VR sickness or VR induced symptoms and effects (VRISE) (Sharples *et al.*, 2008; Bohil *et al.*, 2011; de Franca & Soares, 2017; Palmisano *et al.*, 2017).

Longer durations in a virtual environment have been associated with a higher probability of experiencing VRISE, while the intensity of VRISE also appears to increase proportionally with the duration of the VR session (Sharples *et al.*, 2008). However, extensive linear and angular accelerations provoke intense VRISE, even in a short period of time (McCauley & Sharkey, 1992; LaViola, 2000; Gavgani *et al.*, 2018). VRISE may place the health and safety of the participants or patients at risk of experiencing adverse physiological symptoms (Parsons *et al.*, 2018). Research has also shown that VRISE induce significant decreases in reaction times and overall cognitive performance (Nalivaiko *et al.*, 2015; Nesbitt *et al.*, 2017; Mittelstaedt *et al.*, 2019), as well as substantially increasing body temperatures and heart rates (Nalivaiko *et al.*, 2015), which may compromise physiological data acquisition. Furthermore, the presence of VRISE has been found to significantly augment cerebral blood flow and oxyhemoglobin concentration (Gavgani *et al.*, 2018), electrical brain activity (Arafat *et al.*, 2018), and the connectivity between stimulus-response regions and nausea-processing regions (Toschi *et al.*, 2017). Thus, VRISE appear to confound the reliability of neuropsychological, physiological, and neuroimaging data (Kourtesis *et al.*, 2019).

To our knowledge, there do not appear to be any guidelines as to the appropriate maximum duration of VR research and clinical sessions to evade or alleviate the presence of VRISE. Recently, our work has suggested that VRISE are substantially reduced or prevented by VR software that facilitates ergonomic navigation (e.g., physical movement) and interaction (e.g., direct-hand tracking) facilitated by the hardware capabilities (e.g., motion tracking) of commercial, contemporary VR HMDs comparable to or more advanced than the HTC Vive and/or Oculus Rift (Kourtesis et al., 2019).However, there are other factors such as the type of display and its features that may also induce or reduce VRISE (Mittelstaedt *et al.*, 2018; Kourtesis *et al.*, 2019). Nevertheless, we note that adequate technological competence is required to be able to implement appropriate VR hardware and/or software. In an attempt to reach a methodological consensus, we have proposed minimum hardware and software features, which appraise the suitability of VR hardware and software (see Table 1; Kourtesis *et al.*, 2019).

While VRISE may occur for various reasons, they are predominantly the undesirable outcomes of hardware and software insufficiencies (e.g., low resolution and refresh rates of the image, a narrow field of view, non-ergonomic interactions, and inappropriate navigation modes) (de Franca & Soares, 2017; Palmisano *et al.*, 2017; Kourtesis *et al.*, 2019). In terms of hardware, the technical specifications of the computer (e.g., processing power and graphics card), and VR HMD (e.g., the field of view, refresh rate, and resolution) suffice to appraise their suitability (Kourtesis *et al.*, 2019). However, there is not a tool to quantify the software's recommended features, as well as the intensity of VRISE (Kourtesis *et al.*, 2019). Currently, the most frequently used measure of VRISE is the simulator sickness questionnaire (SSQ), which only considers the symptoms pertinent to simulator sickness (Kennedy *et al.*, 1993). However, the SSQ does not assess software attributes (Kennedy *et al.*, 1993), and there is an argument that simulator sickness symptomatology may not be identical to VRISE (Stanney *et al.*, 1997). There is thus a need for a tool, which will enable researchers to assess both the suitability of VR software, as well as the intensity of VRISE.

Our recent technological literature review of VR hardware and software pinpointed four domains that should be considered in the development or selection of VR research/clinical software (Kourtesis *et al.*, 2019). The domains are user experience, game mechanics, in-game assistance, and VRISE. Each domain has five criteria that should be met to ensure the appropriateness of the software (see Table 1). Also, in the same study, the meta-analysis of 44 VR neuroscientific studies revealed that most of the studies did not report quantitatively VR software's quality and/or VRISE intensity (Kourtesis *et al.*, 2019). In an attempt to provide a brief tool for the appraisal of VR research/clinical software features and VRISE intensity, we developed the virtual reality neuroscience questionnaire (VRNQ), which includes twenty questions that address five criteria under each domain. This study aimed to validate the VRNQ and provide suggestions for the duration of VR research/clinical sessions. We also considered the gaming experience of the participants to examine whether this may affect the duration of the VR sessions. Lastly, we investigated the software predictors of VRISE as measured by the VRNQ.

Table 1. Domains and Criteria for VR Research/Clinical Software

| Domains | User Experience | Game Mechanics | In-Game Assistance | VRISE |
|---|---|---|---|---|
| CRITERIA | An Adequate Level of Immersion | A Suitable Navigation System (e.g., Teleportation) | Digestible Tutorials | Absence or Insignificant Presence of Nausea |
| | Pleasant VR Experience | Availability of Physical Movement | Helpful Tutorials | Absence or Insignificant Presence of Disorientation |
| | High Quality Graphics | Naturalistic Picking/Placing of Items | Adequate Duration of Tutorials | Absence or Insignificant Presence of Dizziness |
| | High Quality Sounds | Naturalistic Use of Items | Helpful In-game Instructions | Absence or Insignificant Presence of Fatigue |
| | Suitable Hardware (HMD and Computer) | Naturalistic 2-Handed Interaction | Helpful In-game Prompts | Absence or Insignificant Presence of Instability |

*Derived from Kourtesis et al. (2019)*

## 2    Methods

### 2.1    Participants

Forty participants (21 males) aged between 28 and 43 years (M = 32.08; SD = 3.54) and an educational level between 12 and 16 full-time years of education (M = 14.25; SD = 1.37) were recruited for the study. Eighteen participants (10 males) identified themselves as gamers through self-report and 22 as non-gamers (11 males). The gamer experience was a dichotomous variable (i.e., gamer or non-gamer) based on the participants' response to a question asking whether they played games on a weekly basis. The participants responded to a call disseminated through mailing lists at the University of Edinburgh and social media. The study was approved by the Philosophy, Psychology and Language Sciences Research Ethics Committee of the University of Edinburgh. All participants provided written informed consent prior to taking part.

### 2.2    Material

#### 2.2.1 Hardware

An HTC Vive HMD with two lighthouse-stations for motion tracking was used with two HTC Vive's wands with 6 degrees of freedom (DoF) to facilitate navigation and interactions within the environment (Kourtesis *et al.*, 2019). The VR area where the participants were immersed and interacted with the virtual environments was 4.4 m2. Additionally, the HMD was connected to a laptop with an Intel Core i7 7700HQ processor at 2.80GHz, 16 GB RAM, a 4095MB NVIDIA GeForce GTX 1070 graphics card, a 931 GB TOSHIBA MQ01ABD100 (SATA) hard disk, and Realtek High Definition Audio.

#### 2.2.2 Software

Three VR games were selected, which included ergonomic navigation (i.e., teleportation and physical mobility) and interactions (i.e., 6 DoF wands simulating hand movements) with the virtual environment. In line with Kourtesis *et al.* (2019), the VR software inclusion criteria (see Table 1) were: 1) ergonomic interactions which simulate real-life hand movements; 2) a navigation system which uses teleportation and physical mobility; 3) comprehensible tutorials pertinent to the controls; and 4) in-game instructions and prompts which assist the user in orientating and interacting with the virtual environment. The suitability of the VR software for both gamers and non-gamers was also considered. The selected VR games which met the above software criteria were: 1) "Job Simulator" (Session 1) [https://store.steampowered.com/app/448280/Job_Simulator/]; 2) "The Lab" (Session 2) [https://store.steampowered.com/app/450390/The_Lab/]; and 3) "Rick and Morty: Virtual Rick-ality" (Session 3) [https://store.steampowered.com/app/469610/Rick_and_Morty_Virtual_Rickality/]. In "Job Simulator", the participant becomes an employee who has several occupations, such as a cook (preparing simply recipes), car mechanic (doing rudimentary tasks e.g., replacing faulty parts), and an office worker (making calls and sending emails). In "The Lab", the participant needs to complete several mini-games like slingshot (shooting down piles of boxes), longbow (shooting down invaders), xortex (spaceship-battles), postcards (visiting exotic places), human medical scan (exploring the human body), solar system (exploring the solar system), robot repair (repairing a robot), and secret shop (exploring a magical shop). In "Rick and Morty: Virtual Rick-ality", the participant needs to complete several imaginary home-chores as in "Job Simulator", though, in this case, the participant is required to follow a sequence of tasks according to a fictional storyline.

### 2.2.3 Virtual Reality Neuroscience Questionnaire (VRNQ)

The VRNQ measures the quality of user experience, game mechanics, and in-game assistance, as well as the intensity of VRISE. The VRNQ involves 20 questions where each question corresponds to one of the criteria for appropriate VR research/clinical software (e.g., the level of immersion; see Table 1). The 20 questions are grouped under four domains, where each domain encompasses five questions. Hence, VNRQ produces a total score corresponding to the overall quality of VR software, as well as four sub-scores (i.e., user experience, game mechanics, in-game assistance, VRISE). The user experience score is based on the intensity of the immersion, the level of enjoyment, as well as the quality of the graphics, sound, and VR technology (i.e., internal and external hardware). The game mechanics' score depends on the ease to navigate, physically move, and interact with the virtual environment (i.e., use, pick & place, and hold items; two-handed interactions). The in-game assistance score appraises the quality of the tutorial(s), in-game instructions (e.g., description of the aim of the task), and prompts (e.g., arrows showing the direction). The VRISE are evaluated by the intensity of primary adverse symptoms and effects pertinent to VR (i.e., nausea, disorientation, dizziness, fatigue, and instability). VRNQ responses are indicated on a 7-point Likert style scale, ranging from 1= extremely low to 7 = extremely high. The higher scores indicate a more positive outcome; this also applies to the evaluation of VRISE intensity. Hence, the higher VRISE score indicates a lower intensity of VRISE (i.e., 1 = extremely intense feeling, 2 = very intense feeling, 3 = intense feeling, 4 = moderate feeling, 5 = mild feeling, 6 = very mild feeling, 7 = absent). The VRNQ also includes space under each question, where the participant may provide optional qualitative feedback. For further details, please see the VRNQ in the supplementary materials.

## 2.3 Procedure

The participants individually attended 3 separate VR sessions; in each session, they were immersed in different VR software. The period between each session was one week for each participant (i.e., 3 weeks in total). The participants went through an induction pertinent to the VR software for that session and the specific HMD and controllers used (i.e., HTC Vive and its 6DoF wands-controllers) before being immersed. Subsequently, the participants were asked to play the respective VR game until they completed it, or they felt any discomfort or fatigue. The duration of each VR session was recorded from the time the software was started until the participant expressed that they wanted to discontinue. At the end of each session, participants were asked to complete the VRNQ. The "Job Simulator" was always used in the 1st session, "The Lab" was always used in the 2nd session, and "Rick and Morty: Virtual Rick-ality" was always used in the 3rd session.

## 2.4 Statistical Analyses

A reliability analysis of the VRNQ was conducted to calculate Cronbach's alpha and inspect whether the items have adequate internal consistency for research and clinical purposes. A Cronbach's alpha of 0.70-1.00 indicates good to excellent internal consistency (Nunally & Bernstein, 1994). A confirmatory factor analysis (CFA) was performed to examine the construct validity of the VRNQ in terms of convergent and discriminant validity (Cole, 1987). The reliability analysis and CFA were conducted using AMOS (version 24) (Arbuckle, 2014), and IBM Statistical Package for the Social Sciences (SPSS) 24.0 (IBM Corp. Released, 2016). Several tests for goodness of fit were implemented to allow the evaluation of VRNQ's structure. The (CFI), Tuckere Lewis index (TLI), standardized root mean square residual (SRMR), and the root mean squared error of approximation (RMSEA) were used to assess model fit. A CFI and TLI equal to or greater than 0.90 indicate good structural model fit to the data (Hu & Bentler, 1999; Jackson *et al.*, 2009; Hopwood & Donnellan, 2010). An SRMR and RMSEA less than 0.08 postulate a good fit to the data (Hu & Bentler, 1999;

Hopwood & Donnellan, 2010). Lastly, the variance of the results was assessed by dividing the χ2 by the degrees of freedom (df), which is an indicator of the sample distribution (Hu & Bentler, 1999; Jackson *et al.*, 2009; Hopwood & Donnellan, 2010).

The reliability and confirmatory factor analyses were conducted based on 120 observations (40 participants * 3 sessions with different software). The a priori sample size calculator for structural equation models was used to calculate the minimum sample size for model structure. This calculator uses the error function formula, the lower bound sample size formula for a structural equation model, and the normal distribution cumulative distribution function (Soper, 2019a), which are in perfect agreement with the recommendations for statistical power analysis for the behavioral sciences (Cohen, 2013). A sample size of 100 observations was suggested as the minimum for conducting CFA to examine the model structure with statistical power equal to or greater than 0.80. Hence, the 120 observations in our sample appear adequate to conduct a CFA with statistical power equal to or greater than 0.80.

Bayesian Pearson correlation analyses were conducted to examine whether any of the demographic variables were significantly associated with the VRNQ total score and sub-scores, or the length of the VR sessions. Bayesian paired samples t-tests were performed to investigate possible differences between each session's duration, as well as the VRNQ results for each VR game. Also, a Bayesian independent samples t-test examined whether there were any differences between gamers and non-gamers in the duration of the session. Lastly, a Bayesian linear regression was performed to examine the predictors of VRISE, where the Jeffreys–Zellner–Siow (JZS) mixed g-prior was used for the selection of the best model. JZS has the computational advantages of a g-prior in conjunction with the theoretical advantages of a Cauchy prior, which are valuable in variable selection for the best model (Liang *et al.*, 2008; Rouder & Morey, 2012). For all the analyses, a Bayes Factor ($BF_{10}$) ≥ 10 was set for statistical inference, which indicates strong evidence in favor of the alternative hypothesis (Rouder & Morey, 2012; Wetzels & Wagenmakers, 2012; Marsman & Wagenmakers, 2017). All the Bayesian analyses were performed using JASP (Version 0.8.1.2) (JASP Team, 2017). The Bayesian Pearson correlation analyses and Bayesian linear regression analysis were conducted based on 120 observations (40 participants * 3 different software sessions). The post-hoc statistical power calculator was used to calculate the observed power of the best model using Bayesian linear regression analysis (Soper, 2019b).

## 3    Results

### 3.1    Reliability Analysis & CFA

The reliability analysis demonstrated good to excellent Cronbach's α for each domain of the VRNQ (i.e., user experience - α = 0.89, game mechanics - α = 0.89, in-game assistance - α = 0.90, VRISE - α = 0.89; see Table 2), which indicate very good internal reliability (Nunally & Bernstein, 1994). VRNQ's fit indices are displayed in Table 2 with their respective thresholds. The χ2/df was 1.61, which indicates good variance in the sample (Hu & Bentler, 1999; Jackson *et al.*, 2009; Hopwood & Donnellan, 2010). Both CFI and TLI were close to 0.95, which suggest a good fit for the VRNQ model (Hu & Bentler, 1999; Jackson *et al.*, 2009; Hopwood & Donnellan, 2010). Comparably, SPMR and RMSEA values were between 0.06 and 0.08, which also support a good fit (Hu & Bentler, 1999; Jackson *et al.*, 2009; Hopwood & Donnellan, 2010). The VRNQ's path diagram is displayed in Figure 1, where from left to right are depicted the correlations among the factors/domains of the VRNQ, the correlations between each factor/domain and its items, and the error terms for each item. The VRNQ items/questions are efficiently associated with their respective factor/domain, which shows good convergent validity (Cole, 1987). Furthermore, there was not any significant correlation amongst the factors/domains, which indicates good discriminant validity (Cole, 1987).

Table 2. Internal Reliability and Goodness of Fit for the VRNQ

| Statistics | Thresholds | Results |
|---|---|---|
| Cronbach's α | ≥ 0.70 | USER – 0.886 |
| | | GM – 0.888 |
| | | GA – 0.895 |
| | | VR – 0.891 |
| $\chi^2$/df | ≤ 2.00 | 1.610 |
| Comparative Fit Index (CFI) | ≥ 0.90 | 0.954 |
| Tuckere Lewis Index (TLI) | ≥ 0.90 | 0.938 |
| Standardised root mean square residual (SRMR) | < 0.08 | 0.076 |
| Root mean square error of approximation (RMSEA) | ≤ 0.08 | 0.071 |

*VRNQ Domains: USER = user experience; GM = game mechanics; GA = in-game assistance; VR = VRISE*

Figure 1. CFA: Model's Path Diagram

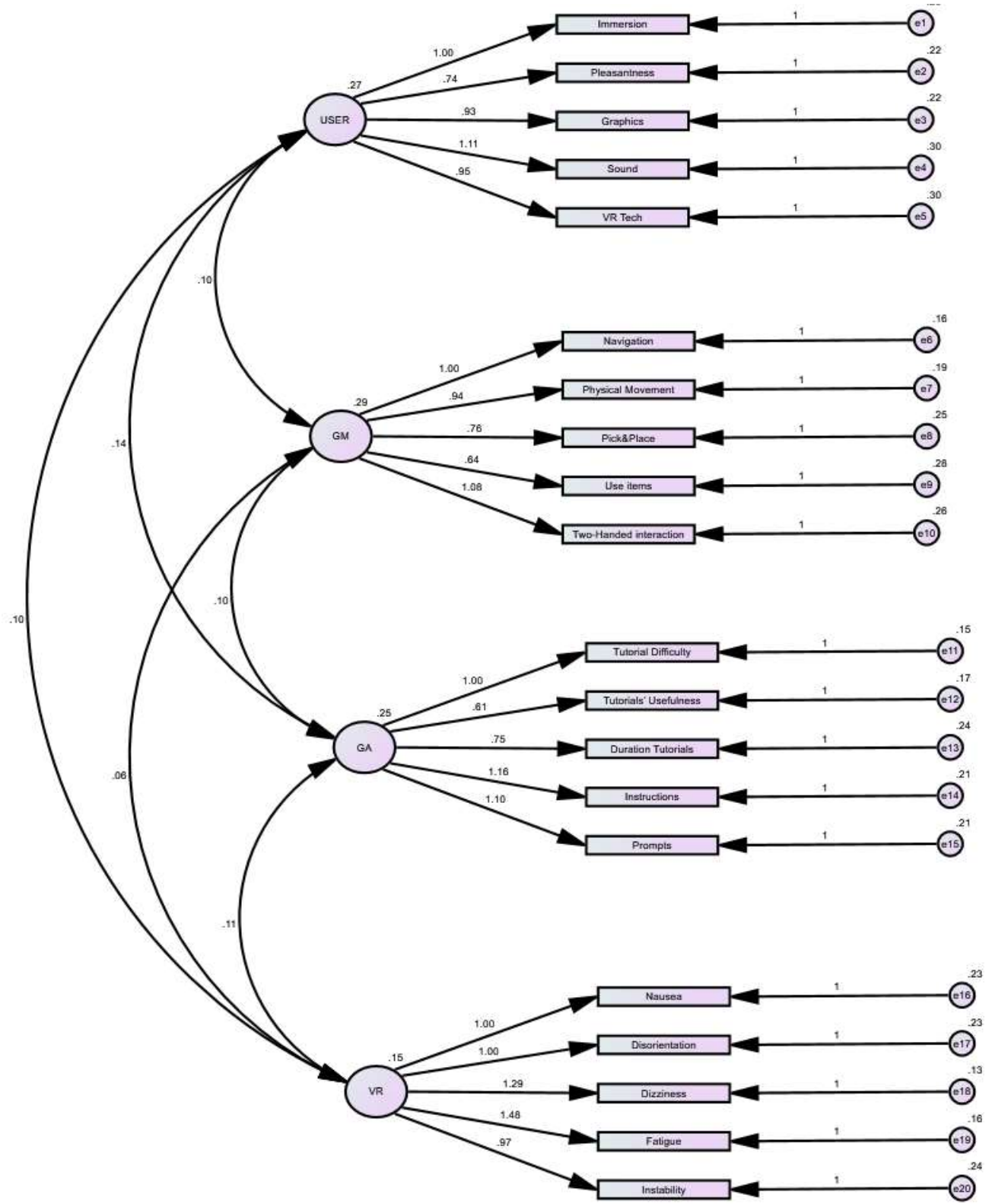

*From left to right: the structural model illustrates the associations between VRNQ domains (paths with double headed arrows) and between each VRNQ domain and its items. At the right there are the error terms (e) for each item; USER = user experience; GM = game mechanics; GA = in-game assistance; VR = VRISE*

## 3.2    Descriptive Statistics of Sessions' Duration and VRNQ Scores

The descriptive statistics for the sessions' durations and the VRNQ scores are displayed in Table 3. In session 1, the participants were immersed for 59.65 (8.42) minutes. In session 1, the average time of gamers seems more than the average time of non-gamers (Table 3). In session 2, the participants spent 64.72 (6.24) minutes (Table 3).  In session 3, gamers spent 70.44 (7.78) minutes, while non-gamers spent 65.73 (6.75) minutes (Table 3). The average total score of the VRNQ for all software was 126.30 (7.55) (maximum score is 140), where gamers and non-gamers scores did not appear to differ. Similarly, the median scores for each domain were 30-32 out of 35, where again gamers and non-gamers scores did not appear to differ. Importantly, all the VRISE scores (per item) for both gamers and non-gamers were equal to 5 (i.e., mild feeling), or 6 (i.e., very mild feeling), or 7 (absent feeling). The vast majority of scores were equal to 6 (i.e., very mild feeling) or 7 (absent feeling) (see Figure 2).

## 3.3    Minimum and Parsimonious Cut-off Scores of VRNQ

Cut-off scores were calculated for the VRNQ total score and sub-scores to inspect the suitability of the assessed VR software (see Table 4). In the VRNQ, the ordinal 1-3 responses are paired with negative qualities, response 4 is paired with neutral/moderate qualities, and 5-7 responses are paired with positive qualities (see Supplementary Material). The minimum cut-offs suggest that if the median of the responses is 25 for every sub-score, and 100 in the total score (i.e., at least a median of 5 for every item), then the VRNQ outcomes indicate that the evaluated VR software is of an adequate quality not to cause any significant VRISE. Furthermore, the parsimonious cut-offs suggest that, if the median of the responses is 30 for every sub-score, and 120 for the total score (i.e., at least a median of 6 for every item) then the utilization of the parsimonious cut-offs more robustly supports the suitability of the VR software. The minimum and parsimonious cut-offs hence appear adequate to guarantee the safety, pleasantness, and appropriateness of the VR software for research and/or clinical purposes.

Table 3. Descriptive Statistics: Duration of VR Sessions and VRNQ Scores

| | Group | N | Mean (SD) | SE |
|---|---|---|---|---|
| **Total Duration** | Gamers | 18 | 199.39 (13.63) | 3.21 |
| | Non-Gamers | 22 | 186.36 (11.76) | 2.51 |
| | Total | 40 | 192.2 (14.09) | 2.23 |
| **Duration of Session 1** | Gamers | 18 | 65.61 (7.14) | 1.68 |
| | Non-Gamers | 22 | 54.77 (5.91) | 1.26 |
| | Total | 40 | 59.65 (8.42) | 1.33 |
| **Duration of Session 2** | Gamers | 18 | 63.33 (6.16) | 1.45 |
| | Non-Gamers | 22 | 65.86 (6.21) | 1.32 |
| | Total | 40 | 64.72 (6.24) | 0.99 |
| **Duration of Session 3** | Gamers | 18 | 70.44 (7.78) | 1.83 |
| | Non-Gamers | 22 | 65.73 (6.75) | 1.44 |
| | Total | 40 | 67.85 (7.52) | 0.69 |
| **VRNQ Total Score Out of 140 (Across 3 Sessions)** | Gamers | 18 | 127.2 (7.32) | 0.99 |
| | Non-Gamers | 22 | 125.6 (7.71) | 0.95 |
| | Total | 40 | 126.3 (7.55) | 0.69 |
| **User's Experience (Across 3 Sessions) Out of 35** | Gamers | 18 | 31.37 (2.73) | 0.34 |
| | Non-Gamers | 22 | 30.91 (2.73) | 0.37 |
| | Total | 40 | 31.12 (2.73) | 0.25 |
| **Game Mechanics (Across 3 Sessions) Out of 35** | Gamers | 18 | 31.50 (2.68) | 0.37 |
| | Non-Gamers | 22 | 31.32 (2.61) | 0.32 |
| | Total | 40 | 31.40 (2.63) | 0.24 |
| **In-Game Assistance (Across 3 Sessions) Out of 35** | Gamers | 18 | 31.70 (2.59) | 0.35 |
| | Non-Gamers | 22 | 31.65 (2.52) | 0.31 |
| | Total | 40 | 31.68 (2.54) | 0.23 |
| **VRISE (Across 3 Sessions) Out of 35** | Gamers | 18 | 32.67 (2.17) | 0.30 |
| | Non-Gamers | 22 | 31.71 (2.56) | 0.32 |
| | Total | 40 | 32.14 (2.43) | 0.22 |

Figure 2. VRISE Intensity in VR Sessions as Measured by VRNQ

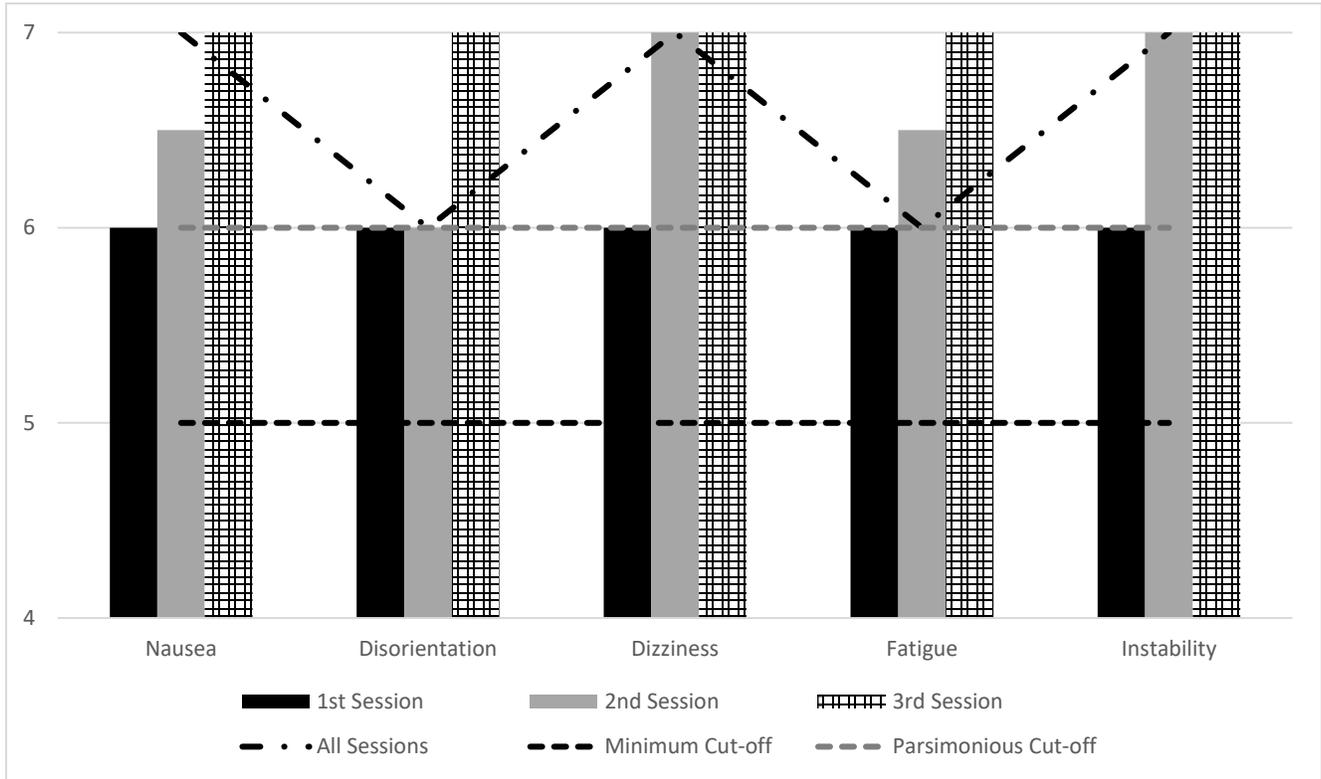

*Median scores of VRISE items of VRNQ; VRNQ Minimum Cut-off (≥); VRNQ Parsimonious Cut-off (≥);
1 = Extreme intense feeling; 2 = Very intense feeling; 3 = Intense feeling; 4 = Moderate feeling;
5 = Mild feeling; 6 = Very mild feeling; 7 = Absent feeling*

Table 4. VRNQ Cut-offs

| Score | Minimum Cut-offs | Parsimonious Cut-offs |
|---|---|---|
| **User Experience** | ≥ 25/35 | ≥ 30/35 |
| **Game Mechanics** | ≥ 25/35 | ≥ 30/35 |
| **In-Game Assistance** | ≥ 25/35 | ≥ 30/35 |
| **VRISE** | ≥ 25/35 | ≥ 30/35 |
| **VRNQ Total Score** | ≥ 100/140 | ≥ 120/140 |

*The median of each sub-score and totals scores should meet the suggested cut-offs to support that the
evaluated VR software has an adequate quality without any significant VRISE. The utilisation of the
parsimonious cut-offs more robustly supports the suitability of the VR software.*

## 3.4   Bayesian T-Tests

The Bayesian independent samples t-test between gamers and non-gamers indicated that the former spent significantly more time in VR across the total duration for the 3 sessions ($BF_{10} = 14.99$), as well as the duration of the 1st session ($BF_{10} = 2,532$; see Table 4) (Wetzels & Wagenmakers, 2012; Marsman & Wagenmakers, 2017). The difference is much smaller in the total duration than the difference in the 1st session. Thus, the difference between the gamers and non-gamers in the total duration appears to be driven by the substantial difference in the 1st session's duration (see Table 5). Conversely, the Bayesian paired samples t-test (i.e., differences between the VR games) indicated significant differences in the total score and every sub-score of VRNQ (see Table 6) between the VR software. The VR software in the 3rd session was evaluated higher than the VR software in the 1st and 2nd sessions, while the VR software in the 2nd session was rated better than the VR software in the 1st session. There was also an important difference between the duration of the 3rd session (longer) and the duration of the 1st session (shorter; $BF_{10} = 103,568$), while there was not a substantial difference between the duration of the 2nd and 3rd sessions ($BF_{10} = 2.78$), as well as between the duration of 1st and 2nd sessions ($BF_{10} = 7.05$; see Table 6) (Wetzels & Wagenmakers, 2012; Marsman & Wagenmakers, 2017).

Table 5. Bayesian Independent Samples T-Test: Gamers against Non-Gamers

| Variables | Significance | $BF_{10}$ | error % |
|---|:---:|:---:|:---:|
| Age | | 0.323 | 0.006 |
| Education | | 0.325 | 0.006 |
| Total Duration | * | 14.987 | 7.044e -6 |
| Session 1 Duration | *** | 2531.886 | 7.491e -8 |
| Session 2 Duration | | 0.595 | 0.006 |
| Session 3 Duration | | 1.580 | 0.003 |
| VRNQ Total | | 0.425 | 0.007 |
| User's Experience | | 0.359 | 0.006 |
| Game Mechanics | | 0.315 | 0.006 |
| In-Game Assistance | | 0.315 | 0.006 |
| VRISE | | 0.745 | 0.003 |

Table 6. Bayesian Paired Samples T-Tests: Differences between the VR Software

| Pairs | | Significance | $BF_{10}$ | error % |
|---|---|---|---|---|
| Session 2 Duration | Session 1 Duration | | 7.049 | ~ 0.001 |
| Session 3 Duration | Session 2 Duration | | 2.783 | ~ 3.276e -4 |
| Session 3 Duration | Session 1 Duration | *** | 103568.858 | NaN |
| S3 VRNQ Total | S2 VRNQ Total | *** | 6.942e +12 | NaN |
| S3 VRNQ Total | S1 VRNQ Total | *** | 3.520e +20 | NaN |
| S2 VRNQ Total | S1 VRNQ Total | *** | 8.500e +17 | NaN |
| S3 VRISE | S2 VRISE | *** | 22075.036 | NaN |
| S3 VRISE | S1 VRISE | *** | 1.322e +10 | NaN |
| S2 VRISE | S1 VRISE | *** | 1.160e +7 | NaN |
| S3 In-Game Assistance | S2 In-Game Assistance | *** | 207216.904 | NaN |
| S2 In-Game Assistance | S1 In-Game Assistance | *** | 1.197e +7 | NaN |
| S3 In-Game Assistance | S1 In-Game Assistance | *** | 8.028e +10 | NaN |
| S3 Game Mechanics | S2 Game Mechanics | *** | 274310.417 | NaN |
| S2 Game Mechanics | S1 Game Mechanics | *** | 4.883e +14 | NaN |
| S3 Game Mechanics | S1 Game Mechanics | *** | 2.876e +14 | NaN |
| S3 User's Experience | S2 User's Experience | *** | 2.873e +7 | NaN |
| S3 In-Game Assistance | S1 User's Experience | *** | 2.597e +7 | NaN |
| S2 User's Experience | S1 User's Experience | *** | 1.708e +6 | NaN |

*$BF_{10}$ = Bayes Factor; \* $BF_{10} > 10$, \*\* $BF_{10} > 30$, \*\*\* $BF_{10} > 100$; S1 = Session 1; S2 = Session 2; S3 = Session 3.*

### 3.5 Bayesian Pearson Correlation Analyses & Regression Analysis

The Bayesian Pearson correlation analyses did not show any significant correlation between age and any of the VRNQ scores, between age and duration of the sessions, between VR education and any of the VRNQ scores, or between education and duration of the sessions. However, the duration of the session was positively correlated with the total VRNQ score ($BF_{10} = 81.54$; $r(120) = 0.310$, $p < .001$). Furthermore, the VRISE score substantially correlated with the following VRNQ items: immersion, pleasantness, graphics, sound, pick & place, tutorial's difficulty, tutorial's usefulness, tutorial's duration, instructions, and prompts (see Table 7). In contrast, VRISE did not significantly correlate with the following VRNQ items: VR tech, navigation, physical movement, use items, or two-handed interactions (see Table 7). Moreover, the Bayesian regression analysis indicated the five best models that predicted the VRNQ's VRISE score (see Table 8). The best model includes the following items from the VRNQ: immersion, graphics, sound, instructions, and prompts. All the predictors exceeded the prior inclusion probabilities (see Figure 3). The best model showed a $BF_M = 117.42$, whereas the second-best model displayed a $BF_M = 56.40$ (see Table 8); hence, the difference between the best model compared to the second-best model was robust (Rouder & Morey,2012; Wetzels & Wagenmakers, 2012; Marsman & Wagenmakers, 2017). Also, the best model has an $R^2 = 0.324$ (see Table 8), which postulates that the model explains the 32.4% of the variance of VRISE score (Rouder & Morey,2012; Wetzels & Wagenmakers, 2012). Lastly, the post-hoc statistical power analysis for the best model indicated an observed statistical power of 0.998, p <.001, which postulates a high efficiency, precision, reproducibility, and reliability of the regression analysis and results (Button *et al.*, 2013; Cohen, 2013).

Table 7. Bayesian Pearson Correlations Analyses: VRISE Score with VRNQ Items

| Pairs | | Significance | BF$_{10}$ | r |
|---|---|---|---|---|
| VRISE | Immersion | *** | 1226.538 | 0.371 |
| VRISE | Pleasantness | * | 20.504 | 0.273 |
| VRISE | Graphics | *** | 1629.195 | 0.377 |
| VRISE | Sound | *** | 18586.578 | 0.421 |
| VRISE | VR Tech | | 5.094 | 0.228 |
| VRISE | Navigation | | 4.808 | 0.226 |
| VRISE | Physical Movement | | 2.229 | 0.197 |
| VRISE | Pick & Place | *** | 175.087 | 0.329 |
| VRISE | Use Items | | 0.405 | 0.109 |
| VRISE | Two-Handed Interaction | | 0.506 | 0.123 |
| VRISE | Tutorial Difficulty | *** | 28252.587 | 0.428 |
| VRISE | Tutorials Usefulness | *** | 161.949 | 0.327 |
| VRISE | Tutorials' Duration | *** | 128.539 | 0.322 |
| VRISE | Instructions | *** | 952.871 | 0.366 |
| VRISE | Prompts | *** | 706510.726 | 0.476 |

*BF$_{10}$ = Bayes Factor; \* BF$_{10}$ > 10, \*\* BF$_{10}$ > 30, \*\*\* BF$_{10}$ > 100;*

Table 8. Models' Comparison: Predictors of VRISE score

| Models | P(M) | P(M\|data) | BF$_M$ | BF$_{10}$ | R$^2$ |
|---|---|---|---|---|---|
| Prompts + Sound + Graphics + Immersion + Instructions | 0.004 | 0.304 | 117.42 *** | 1.000 | 0.324 |
| Prompts + Graphics + Immersion + Instructions + Pleasantness | 0.004 | 0.173 | 56.47 ** | 0.571 | 0.317 |
| Prompts + Sound + Graphics + Immersion + Instructions + Pick & Place | 0.004 | 0.161 | 43.15 * | 0.443 | 0.330 |
| Prompts + Sound + Graphics + Immersion + Instructions + Pick &Place + Tutorials Usefulness + Pleasantness | 0.021 | 0.123 | 6.62 | 0.072 | 0.337 |
| Prompts + Graphics + Immersion + Instructions + Pick & Place + Tutorials Usefulness + Pleasantness | 0.008 | 0.077 | 10.72 * | 0.121 | 0.329 |

*P = Probability; M = Model; BF$_M$ = Model's Bayesian Factor; * BF$_M$ >10, ** BF$_M$ >30, *** BF$_M$ >100; BF$_{10}$ = BF against null model*

Figure 3. Variables' Prior Inclusion Probabilities

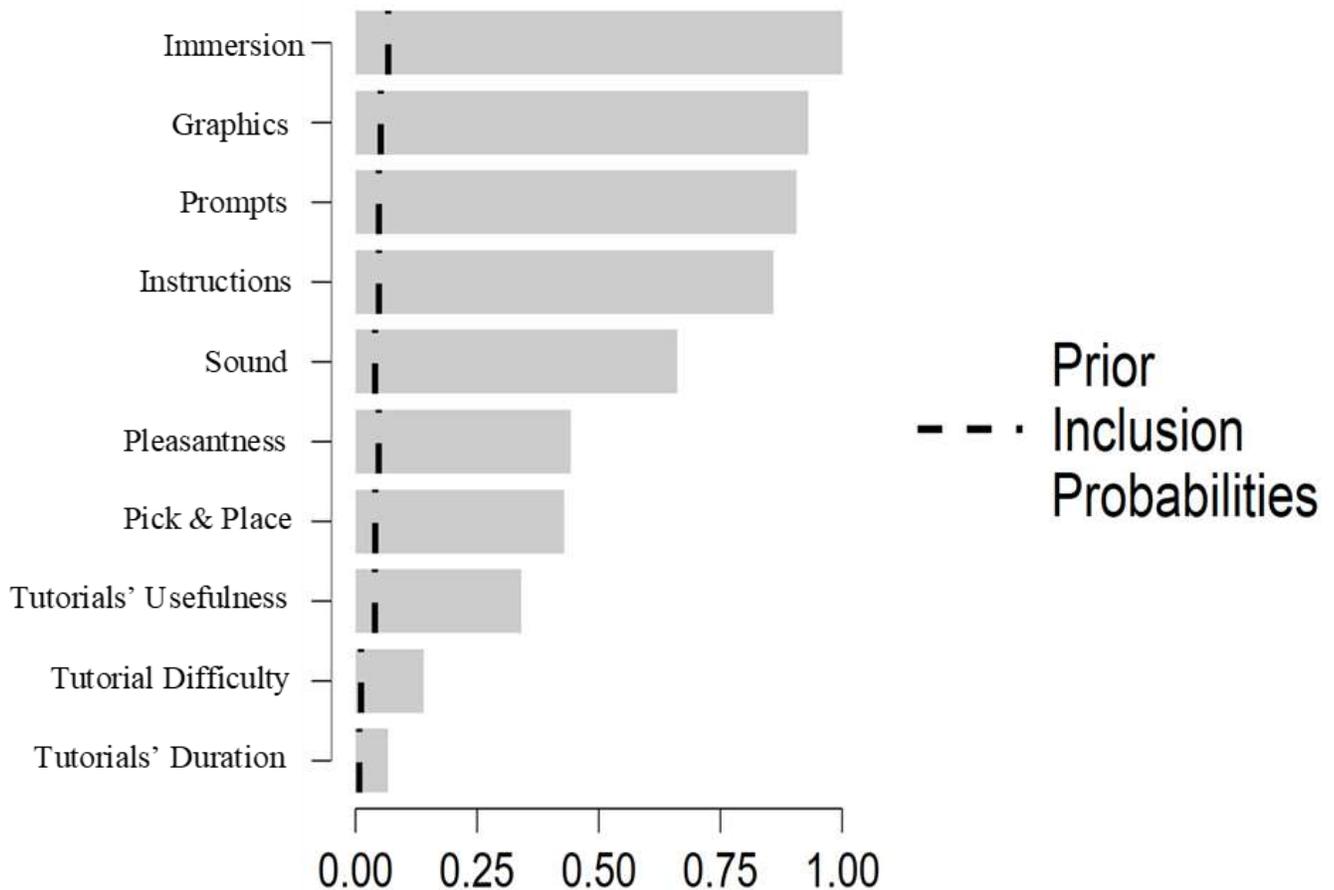

## 4    Discussion

### 4.1    The VRNQ as a Research and Clinical Tool

The VRNQ is a short questionnaire (5-10 minutes administration time) which assesses the quality of VR software in terms of user experience, game mechanics, in-game assistance, and VRISE. The values of the fit indices of CFA (i.e., CFI, TLI, SPMR and RMSEA) indicated that the VRNQ's structure was a good fit to the data, which postulates good construct validity for the VRNQ (Hu & Bentler, 1999; Jackson *et al.*, 2009; Hopwood & Donnellan, 2010). In addition, the construct validity of the VRNQ was supported by its convergent and discriminant validity (Cole, 1987). VRNQ items were strongly correlated with their grouping factor, which indicates robust convergent validity, while there were substantially poor correlations between the factors, which postulates very good discriminant validity (Cole, 1987). Furthermore, the Cronbach's α for each VRNQ domain (i.e., user experience - α = 0.89, game mechanics - α = 0.89, in-game assistance - α = 0.90, VRISE - α = 0.89; see Table 2) suggest very good construct validity (Nunally & Bernstein, 1994). Henceforth, the VRNQ emerges as a valid and suitable tool to evaluate the quality of the VR research/clinical software as well as the intensity of the adverse VRISE.

Furthermore, minimum and parsimonious cut-off scores were calculated for the VRNQ total score and sub-scores to inspect the suitability of the assessed VR software. The minimum cut-offs indicate the lowest acceptable quality that VR research/clinical software should be, while the parsimonious cut-offs are offered for more robust support of the VR software's suitability, which may be required in experimental and clinical designs with more conservative standards. However, the individual scores from the VRNQ may be modulated by individual differences and preferences unrelated to the quality of the software (Kortum & Peres, 2014). In addition, the VRNQ produces ordinal data; therefore, the median is the appropriate measure for their analysis (Harpe, 2015). Hence, the median VRNQ scores for the whole sample should be used to assess the VR software's quality effectively. Also, the medians of the VRNQ total score and sub-scores allow the generalisation of the results and comparison between different VR software (Kortum & Peres, 2014; Harpe, 2015). Researchers, clinicians, and/or research software developers should use the medians of the VRNQ total score and sub-scores to assess whether the implemented VR software exceed the minimum or parsimonious cut-offs. Hence, if the medians of the VRNQ sub-scores and totals score for VR research software meet the minimum cut-offs, then these results support the VR software's suitability. Likewise, if the medians of VRNQ sub-scores and totals score for VR research software meet the parsimonious cut-offs, then these results provide even stronger support for its suitability. However, median scores below these cut-offs suggest that the suitability of the VR software is questionable, but they do not indicate that this VR software is certainly unsuitable.

 Also, VRNQ appears as an appropriate tool to measure both VRISE and VR software features compared to other questionnaires. The SSQ is the most implemented questionnaire in VR studies. However, the SSQ only considers the symptoms pertinent to simulator sickness and it does not assess software attributes (Kennedy *et al.*, 1993), while there is a dispute that simulator sickness symptomatology may not be the same as VRISE (Stanney *et al.*, 1997). Alternatively, Virtual reality sickness questionnaire (VRSQ) was recently developed (Kim *et al.*, 2018). The development of VRSQ was based on the SSQ, where the researchers attempted to isolate the items which are pertinent to VRISE (Kim *et al.*, 2018).  However, their sample size was relatively small (i.e., 24 participants * 4 sessions = 96 observations) (Kim *et al.*, 2018).  Notably, the factor analyses of Kim *et al.*, (2018) accepted only items pertinent to oculomotor and disorientation components of SSQ, and rejected all the items pertinent to nausea (i.e.,7 items) (Kim *et al.*, 2018), while nausea is the most frequent symptom in VRISE (Stanney *et al.*, 1997; Sharples *et al.*, 2008; Bohil *et al.*, 2011; de Franca & Soares, 2017; Palmisano *et al.*, 2017). Also, comparable to SSQ, VRSQ does not consider software features. Hence, the VRNQ appears to be the only valid and suitable tool to evaluate both the intensity of predominant VRISE and the quality of VR software features.

The VRNQ allows researchers to report the quality of VR software and/or the intensity of VRISE in their VR studies. However, an in-depth assessment of the numerous software features requires a questionnaire with more than the 20 questions of the VRNQ (Zarour *et al.*, 2015). For an in-depth software analysis, questionnaires with more questions pertinent to the whole spectrum of software features should be preferred (Zarour *et al.*, 2015). Additionally, the VRNQ has solely five items pertinent to VRISE. Hence, it does not offer an exhaustive assessment of VRISE. Studies that aim to investigate VRISE in depth should opt for a tool which contains more items pertinent to VRISE than VRNQ (e.g., SSQ). The VNRQ is a brief questionnaire (5-10 minutes administration time) including 20 items, which enables researchers, clinicians, and research software developers to evaluate and report the quality of the VR software and the intensity of VRISE for research and clinical purposes.

## 4.2   Maximum Duration of VR Sessions

The duration of the VR session is a crucial factor in research and/or clinical design. In our sample, the participants discontinued the VR session due to loss of interest, while none discontinued due to VRISE. In the 1st session, gamers spent significantly more time immersed than the non-gamers; a difference which modulated the difference between the two groups in the summed duration across all sessions. However, it is worth noting that there was not a significant difference between the two groups in the time spent in VR for the 2nd and 3rd sessions. The observed difference in the 1st session and the absence of a difference in the later sessions' durations postulates that when users are familiarized with the VR technology, while the influence of their gaming experience on the session's duration becomes insignificant. In support of this, a recent study showed that user gaming experience does not affect the perceived workload of the users in VR (Lum *et al.*, 2018). Hence, the level of familiarization of the participants with the VR technology appears to affect substantially the duration of the VR session.

Nevertheless, in the whole sample, irrespective of participants' gaming experience, the durations of the 2nd and 3rd sessions are sufficiently longer than the duration of the 1st session. The duration of the 3rd session is not significantly longer than the duration of the 2nd session. Furthermore, given that in each session, a different VR software was administered, the VRNQ correspondingly pinpointed significant differences amongst the implemented VR software' quality. All the VRNQ scores for the 3rd session's VR software are greater than the 2nd session's VR software scores. Similarly, all the VRNQ scores for the 2nd session's VR software are greater than the 1st session's VR software scores. Also, the duration of VR session was positively correlated with the total score of VRNQ. Thus, the quality of the VR software as measured by the VRNQ seems to be significantly associated with the duration of the VR session.

Overall, in every session, the intensity of VRISE was reported as very mild to absent by the vast majority of the sample. However, comparable to the rest of the VRNQ scores, the VRISE score for the 3rd VR session was significantly higher (i.e., milder feeling) than the 2nd and 3rd sessions. Similarly, the VRISE score for the 2nd session's VR software was substantially higher than the 1st session's VR software score. Notably, there was not any difference between gamers and non-gamers in the VRNQ scores across the three sessions. Equally, the age and education of participants did not correlate with any of the VRNQ scores or the duration of sessions. Thus, the age, education, and gaming experience of the participants did not affect the responses in the VRNQ. Therefore, the observed differences in the VRISE scores between the VR sessions support that the quality of the VR software as measured by the VRNQ and the level of familiarization of the participants with the VR technology also affect the intensity of VRISE.

The findings postulate that the implementation of VR software with a maximum duration between 55-70 minutes is substantially feasible. However, long exposures in VR have been found to increase the probability of experiencing VRISE and the intensity of VRISE (Sharples *et al.*, 2008). In our sample, especially in the 3rd session, which was substantially longer than the other sessions, the intensity of VRISE was significantly lower than the rest of the sessions. As discussed above, the substantially lower intensity of VRISE in the 3rd session appears to be a result of increased VR familiarity, and the better quality of the implemented VR software as measured by the VRNQ. Hence, researchers and/or clinicians should consider the quality of their VR software to define the appropriate duration of their VR session. In research and clinical designs where the duration of the VR session is required to be between 55-70 minutes, the researchers and/or clinicians should opt for the parsimonious cut-offs of the VRNQ to ensure adequate quality of their VR software to facilitate longer sessions without significant VRISE. Additionally, an extended introductory tutorial which allows participants to familiarize themselves with the VR technology and mechanics would assist with the implementation of longer (i.e., 55-70 minutes) VR sessions, where the presence and intensity of VRISE would not be significant.

## 4.3    The Quality of VR Software and VRISE

The VRISE score substantially correlated with almost every item under the section of user experience and in-game assistance (see Table 6). However, the VRISE score did not correlate with VR tech (the item under the user experience's domain) or most of the items under the section of game mechanics. The quality of VR hardware (i.e., the HMD and its controllers) and interactions (i.e., ergonomic or non-ergonomic) with the virtual environment are crucial for the alleviation or evasion of VRISE (Kourtesis *et al.*, 2019). Nevertheless, in this sample, the VR tech item (i.e., the quality of the internal and external VR hardware) was not expected to correlate with the VRISE score, because the HMD and its 6DoF controllers were the same for all 3 VR software versions and sessions. Hence, the variance in the responses to this item was limited. Also, the three VR software games share common game mechanics, especially the same navigation system (i.e., teleportation) and a similar amount of physical mobility. Likewise, apart from some controls (i.e., the button to grab items), the interaction systems of the implemented VR software were very proximal. Therefore, the absence of a correlation between VRISE scores and most of the items in the game mechanics' section was also an expected outcome. Nonetheless, the VRISE score was strongly associated with the level of immersion and enjoyment, the quality of graphics and sound, the comfort to pick and place 3D objects, and the usefulness of in-game assistance modes (i.e., tutorials, instructions, and prompts).

The items which correlated with the VRISE score were also included in the best models of predicting its value (see Table 7). Importantly, the best model includes as predictors of VRISE, the level of immersion, the quality of graphics and sound, and the helpfulness of in-game instructions and prompts (see Table 7). The higher scores for prompts and instructions indicate that the user was substantially assisted by the in-game assistance (e.g., an arrow showing the direction that the user should follow) to orientate and guide his or herself from one point of interest to the next in accordance with the scenario of the VR experience. This may be interpreted as ease to orient and interact with the virtual environment, as well as a significant decrease in confusion (Brade *et al.*, 2018). The quality of the in-game assistance methods is essential for the usability and enjoyment that VR software offers (Brade *et al.*, 2018). Equally, the quality of the graphics is predominantly dependent upon rendering which encompasses the in-game quality of the image known as perceptual quality, and the exclusion of redundant visual information known as occlusion culling (Lavoué & Mantiuk, 2015). The improvement of these two factors not only results in improved quality of the graphics but also in improved performance of the software (Brennesholtz, 2018). Furthermore, the spatialized sound of VR software, which assists the user to orient his or herself (Ferrand *et al.*, 2017), deepens the experienced immersion (Riecke *et al.*, 2011), and enriches the geometry of the virtual space without affecting the performance of the software (Kobayashi *et al.*, 2015). Lastly, the level of immersion appears to be negatively correlated with the frequency and intensity of VRISE (Milleville-Pennel & Charron, 2015; Weech *et al.*, 2019). The best model hence aligns with the relevant literature and provides further evidence in support of the utility of the VRNQ as a valid and efficient tool to appraise the quality of the VR software and intensity of VRISE.

## 4.4   Limitations and Future Studies

This study also has some limitations. In this study, construct validity for the VRNQ is provided. However, future work should endeavor to provide convergent validation of the VRNQ with tools that measure VRISE symptomatology (e.g., SSQ) and/or VR software attributes. Moreover, the sample size was relatively small, but it offered an adequate statistical power for the conducted analyses. Also, the VRNQ does not directly quantify linear or angular accelerations, which may induce intense VRISE in a relatively short period of time (McCauley & Sharkey, 1992; LaViola, 2000; Gavgani *et al.*, 2018). However, the VRNQ quantifies the effect(s) of linear and angular accelerations (i.e., VRISE), where VR software with a highly provocative content (e.g., linear and angular accelerations) would fail to meet or exceed the VRNQ cut-offs for the VRISE domain. Furthermore, the study utilized only one type of VR hardware, which did not allow us to inspect the effect of VR HMD's quality on VRISE presence and intensity. Similarly, our VR software did not allow us to compare different ergonomic interactions or levels of provocative potency pertaining to VRISE. Future studies with a larger sample, various types of VR hardware, and VR software with substantially more diverse features will offer further insights on the impact of software features on VRISE intensity, as well as provide additional support for the VRNQ's structural model. Lastly, neuroimaging (e.g., electroencephalography) and physiological data (e.g., heart rates) may correlate, classify, and predict VRISE symptomatology (Kim *et al.*, 2005; Dennison *et al.*, 2016; Dennison *et al.*, 2019). Hence, future studies should consider collecting neuroimaging and/or physiological data that could further elucidate the relationship between VRNQ's VRISE score(s) and brain region activation or cardiovascular responses (e.g., heart rate).

## 4.5   Conclusion

This study showed that the VRNQ is a valid and reliable tool which assesses the quality of VR software and intensity of VRISE. Our findings support the viability of VR sessions with a duration up to 70 minutes, when the participants are familiarized with VR tech through an induction session, and the quality of the VR software meets the parsimonious cut-offs of VRNQ. Also, our results offered insights on the software-related predictors of VRISE intensity, such as the level of immersion, the quality of graphics and sound, and the helpfulness of in-game instructions and prompts. Finally, the VRNQ enables researchers to quantitatively assess and report the quality of VR software features and intensity of VRISE, which are vital for the efficacious implementation of immersive VR systems in cognitive neuroscience and neuropsychology. The minimum and parsimonious cut-offs of VRNQ may appraise the suitability of VR software for implementation in research and clinical settings. The VRNQ and the findings of this study contribute to the endeavor of establishing thorough VR research and clinical methods that are crucial to guarantee the viability of implementing immersive VR systems in cognitive neuroscience and neuropsychology.

## 5    Conflict of Interest

The authors declare that the research was conducted in the absence of any commercial or financial relationships that could be construed as a potential conflict of interest.

## 6    Author Contributions

The primary author had the initial idea and contributed to every aspect of this study. The rest of the authors contributed to the methodological aspects and the discussion of the results. The VRNQ may be downloaded from supplementary material.

Table 2. Domains and Criteria for VR Research/Clinical Software

| Domains | User Experience | Game Mechanics | In-Game Assistance | VRISE |
|---|---|---|---|---|
| | An Adequate Level of Immersion | A Suitable Navigation System (e.g., Teleportation) | Digestible Tutorials | Absence or Insignificant Presence of Nausea |
| | Pleasant VR Experience | Availability of Physical Movement | Helpful Tutorials | Absence or Insignificant Presence of Disorientation |
| CRITERIA | High Quality Graphics | Naturalistic Picking/Placing of Items | Adequate Duration of Tutorials | Absence or Insignificant Presence of Dizziness |
| | High Quality Sounds | Naturalistic Use of Items | Helpful In-game Instructions | Absence or Insignificant Presence of Fatigue |
| | Suitable Hardware (HMD and Computer) | Naturalistic 2-Handed Interaction | Helpful In-game Prompts | Absence or Insignificant Presence of Instability |

*Derived from Kourtesis et al. (2019)*

Table 3. Domains and Criteria for VR Research/Clinical Software

| *Domains* | **User Experience** | **Game Mechanics** | **In-Game Assistance** | **VRISE** |
|---|---|---|---|---|
| | An Adequate Level of Immersion | A Suitable Navigation System (e.g., Teleportation) | Comprehensible Tutorials | Absence or Insignificant Presence of Nausea |
| | Pleasant VR Experience | Availability of Physical Movement | Helpful Tutorials | Absence or Insignificant Presence of Disorientation |
| *Criteria* | High-Quality Graphics | Naturalistic Picking/Placing of Items | Adequate Duration of Tutorials | Absence or Insignificant Presence of Dizziness |
| | High-Quality Sounds | Naturalistic Use of Items | Helpful In-game Instructions | Absence or Insignificant Presence of Fatigue |
| | Suitable Hardware (HMD and Computer) | Naturalistic 2-Handed Interaction | Helpful In-game Prompts | Absence or Insignificant Presence of Instability |

*Derived from Kourtesis et al., 2019*

Table 2. Internal Reliability and Goodness of Fit for the VRNQ

| Statistics | Thresholds | Results |
|---|---|---|
| Cronbach's $\alpha$ | $\geq 0.70$ | 0.889 |
| $\chi^2$/df | $\leq 2.00$ | 1.610 |
| Comparative Fit Index (CFI) | $\geq 0.90$ | 0.954 |
| Tuckere Lewis Index (TLI) | $\geq 0.90$ | 0.938 |
| Standardised root mean square residual (SRMR) | $< 0.08$ | 0.076 |
| Root mean square error of approximation (RMSEA) | $\leq 0.08$ | 0.071 |

Table 3. Descriptive Statistics: Duration of VR Sessions and VRNQ Scores

| | | | | |
|---|---|---|---|---|
| **Total Duration** | Gamers | 18 | 199.39 (13.63) | 3.21 |
| | Non-Gamers | 22 | 186.36 (11.76) | 2.51 |
| | Total | 40 | 192.2 (14.09) | 2.23 |
| **Duration of Session 1** | Gamers | 18 | 65.61 (7.14) | 1.68 |
| | Non-Gamers | 22 | 54.77 (5.91) | 1.26 |
| | Total | 40 | 59.65 (8.42) | 1.33 |
| **Duration of Session 2** | Gamers | 18 | 63.33 (6.16) | 1.45 |
| | Non-Gamers | 22 | 65.86 (6.21) | 1.32 |
| | Total | 40 | 64.72 (6.24) | 0.99 |
| **Duration of Session 3** | Gamers | 18 | 70.44 (7.78) | 1.83 |
| | Non-Gamers | 22 | 65.73 (6.75) | 1.44 |
| | Total | 40 | 67.85 (7.52) | 0.69 |
| **VRNQ Total Score Out of 140 (Across 3 Sessions)** | Gamers | 18 | 127.2 (7.32) | 0.99 |
| | Non-Gamers | 22 | 125.6 (7.71) | 0.95 |
| | Total | 40 | 126.3 (7.55) | 0.69 |
| **User's Experience (Across 3 Sessions) Out of 35** | Gamers | 18 | 31.37 (2.73) | 0.34 |
| | Non-Gamers | 22 | 30.91 (2.73) | 0.37 |
| | Total | 40 | 31.12 (2.73) | 0.25 |
| **Game Mechanics (Across 3 Sessions) Out of 35** | Gamers | 18 | 31.50 (2.68) | 0.37 |
| | Non-Gamers | 22 | 31.32 (2.61) | 0.32 |
| | Total | 40 | 31.40 (2.63) | 0.24 |
| **In-Game Assistance (Across 3 Sessions) Out of 35** | Gamers | 18 | 31.70 (2.59) | 0.35 |
| | Non-Gamers | 22 | 31.65 (2.52) | 0.31 |
| | Total | 40 | 31.68 (2.54) | 0.23 |
| **VRISE (Across 3 Sessions) Out of 35** | Gamers | 18 | 32.67 (2.17) | 0.30 |
| | Non-Gamers | 22 | 31.71 (2.56) | 0.32 |
| | Total | 40 | 32.14 (2.43) | 0.22 |

Table 4. VRNQ Cut-offs

| Score | Minimum Cut-offs | Parsimonious Cut-offs |
|---|---|---|
| **User Experience** | ≥ 25/35 | ≥ 30/35 |
| **Game Mechanics** | ≥ 25/35 | ≥ 30/35 |
| **In-Game Assistance** | ≥ 25/35 | ≥ 30/35 |
| **VRISE** | ≥ 25/35 | ≥ 30/35 |
| **VRNQ Total Score** | ≥ 100/140 | ≥ 120/140 |

*The median of each sub-score and totals scores should meet the suggested cut-offs to support that the evaluated VR software has an adequate quality without any significant VRISE. The utilisation of the parsimonious cut-offs more robustly supports the suitability of the VR software.*

Table 5. Bayesian Independent Samples T-Test: Gamers against Non-Gamers

| Variables | Significance | $BF_{10}$ | error % |
|---|---|---|---|
| Age | | 0.323 | 0.006 |
| Education | | 0.325 | 0.006 |
| Total Duration | * | 14.987 | 7.044e -6 |
| Session 1 Duration | *** | 2531.886 | 7.491e -8 |
| Session 2 Duration | | 0.595 | 0.006 |
| Session 3 Duration | | 1.580 | 0.003 |
| VRNQ Total | | 0.425 | 0.007 |
| User's Experience | | 0.359 | 0.006 |
| Game Mechanics | | 0.315 | 0.006 |
| In-Game Assistance | | 0.315 | 0.006 |
| VRISE | | 0.745 | 0.003 |

Table 6. Bayesian Paired Samples T-Tests: Differences between the VR Software

| Pairs | | Significance | BF$_{10}$ | error % |
|---|---|---|---|---|
| Session 2 Duration | Session 1 Duration | | 7.049 | ~ 0.001 |
| Session 3 Duration | Session 2 Duration | | 2.783 | ~ 3.276e -4 |
| Session 3 Duration | Session 1 Duration | *** | 103568.858 | NaN |
| S3 VRNQ Total | S2 VRNQ Total | *** | 6.942e +12 | NaN |
| S3 VRNQ Total | S1 VRNQ Total | *** | 3.520e +20 | NaN |
| S2 VRNQ Total | S1 VRNQ Total | *** | 8.500e +17 | NaN |
| S3 VRISE | S2 VRISE | *** | 22075.036 | NaN |
| S3 VRISE | S1 VRISE | *** | 1.322e +10 | NaN |
| S2 VRISE | S1 VRISE | *** | 1.160e +7 | NaN |
| S3 In-Game Assistance | S2 In-Game Assistance | *** | 207216.904 | NaN |
| S2 In-Game Assistance | S1 In-Game Assistance | *** | 1.197e +7 | NaN |
| S3 In-Game Assistance | S1 In-Game Assistance | *** | 8.028e +10 | NaN |
| S3 Game Mechanics | S2 Game Mechanics | *** | 274310.417 | NaN |
| S2 Game Mechanics | S1 Game Mechanics | *** | 4.883e +14 | NaN |
| S3 Game Mechanics | S1 Game Mechanics | *** | 2.876e +14 | NaN |
| S3 User's Experience | S2 User's Experience | *** | 2.873e +7 | NaN |
| S3 In-Game Assistance | S1 User's Experience | *** | 2.597e +7 | NaN |
| S2 User's Experience | S1 User's Experience | *** | 1.708e +6 | NaN |

*BF$_{10}$ = Bayes Factor; * BF$_{10}$ > 10, ** BF$_{10}$ > 30, *** BF$_{10}$ > 100; S1 = Session 1; S2 = Session 2;*

*S3 = Session 3.*

Table 7. Bayesian Pearson Correlations Analyses: VRISE Score with VRNQ Items

| Pairs | | Significance | BF$_{10}$ | r |
|---|---|---|---|---|
| VRISE | Immersion | *** | 1226.538 | 0.371 |
| VRISE | Pleasantness | * | 20.504 | 0.273 |
| VRISE | Graphics | *** | 1629.195 | 0.377 |
| VRISE | Sound | *** | 18586.578 | 0.421 |
| VRISE | VR Tech | | 5.094 | 0.228 |
| VRISE | Navigation | | 4.808 | 0.226 |
| VRISE | Physical Movement | | 2.229 | 0.197 |
| VRISE | Pick & Place | *** | 175.087 | 0.329 |
| VRISE | Use Items | | 0.405 | 0.109 |
| VRISE | Two-Handed Interaction | | 0.506 | 0.123 |
| VRISE | Tutorial Difficulty | *** | 28252.587 | 0.428 |
| VRISE | Tutorials Usefulness | *** | 161.949 | 0.327 |
| VRISE | Tutorials' Duration | *** | 128.539 | 0.322 |
| VRISE | Instructions | *** | 952.871 | 0.366 |
| VRISE | Prompts | *** | 706510.726 | 0.476 |

*BF$_{10}$ = Bayes Factor; \* BF$_{10}$ > 10, \*\* BF$_{10}$ > 30, \*\*\* BF$_{10}$ > 100;*

Table 8. Models' Comparison: Predictors of VRISE score

| Models | P(M) | P(M\|data) | BF$_M$ | BF$_{10}$ | R$^2$ |
|---|---|---|---|---|---|
| Prompts + Sound + Graphics + Immersion + Instructions | 0.004 | 0.304 | 117.42 *** | 1.000 | 0.324 |
| Prompts + Graphics + Immersion + Instructions + Pleasantness | 0.004 | 0.173 | 56.47 ** | 0.571 | 0.317 |
| Prompts + Sound + Graphics + Immersion + Instructions + Pick & Place | 0.004 | 0.161 | 43.15 * | 0.443 | 0.330 |
| Prompts + Sound + Graphics + Immersion + Instructions + Pick &Place + Tutorials Usefulness + Pleasantness | 0.021 | 0.123 | 6.62 | 0.072 | 0.337 |
| Prompts + Graphics + Immersion + Instructions + Pick & Place + Tutorials Usefulness + Pleasantness | 0.008 | 0.077 | 10.72 * | 0.121 | 0.329 |

*P = Probability; M = Model; BF$_M$ = Model's Bayesian Factor; \* BF$_M$ >10, \*\* BF$_M$ >30, \*\*\* BF$_M$ >100;*
*BF$_{10}$ = BF against null model*